\documentclass[aps,prd,preprint,superscriptaddress,showpacs]{revtex4}

\usepackage{amsfonts}
\usepackage{mathrsfs}
\usepackage{amssymb}
\usepackage{amsmath}
\usepackage{graphicx,epsfig}

\def\bwt{\begin{widetext}}

\def\ewt{\end{widetext}}

\def\be{\begin{equation}}

\def\ee{\end{equation}}

\def\bea{\begin{eqnarray}}

\def\eea{\end{eqnarray}}

\def\bean{\begin{eqnarray*}}

\def\eean{\end{eqnarray*}}

\def\bary{\begin{array}}

\def\eary{\end{array}}

\def\bit{\begin{itemize}}

\def\eit{\end{itemize}}

\def\su5u1{SU(5) \times U(1)}

\def\fsu5u1{SU(5) \times U(1)'}

\def\so10{SO(10)}

\def\sq20{SO(10) \times SO(10)}

\usepackage{amssymb}

\begin{document}

\setlength{\parskip}{0cm}

\title{ Generalizing Minimal Supergravity }

\author{Tianjun Li}

\affiliation{George P. and Cynthia W. Mitchell Institute for
Fundamental Physics, Texas A$\&$M University, College Station, TX
77843, USA }

\affiliation{Key Laboratory of Frontiers in Theoretical Physics,
      Institute of Theoretical Physics, Chinese Academy of Sciences,
Beijing 100190, P. R. China }

\author{Dimitri V. Nanopoulos}

\affiliation{George P. and Cynthia W. Mitchell Institute for
Fundamental Physics,
 Texas A$\&$M University, College Station, TX 77843, USA }

\affiliation{Astroparticle Physics Group,
Houston Advanced Research Center (HARC),
Mitchell Campus, Woodlands, TX 77381, USA}

\affiliation{Academy of Athens, Division of Natural Sciences,
 28 Panepistimiou Avenue, Athens 10679, Greece }

%\date{\today}

%%%%%%%%%%%%%%%%%%%%%%%%%%%%%%%%%%%%%%%%%%%%%%%%%%%%%%%%%%%%%%%%%%%%%%%%%%%%

\begin{abstract}

In Grand Unified Theories (GUTs),
 the Standard Model (SM) gauge couplings need not be unified
at the GUT scale due to the high-dimensional operators. 
Considering gravity mediated supersymmetry breaking,
we study for the first time the generic gauge coupling relations at
the GUT scale, and  the general gaugino mass relations 
which are valid from the GUT scale to the electroweak scale 
at one loop. We define the index $k$ for these relations, 
which can be calculated in GUTs and can be determined at 
the Large Hadron Collider and the future International Linear 
Collider. Thus, we give a concrete definition of the GUT scale
in these theories, and suggest a new way to test  general GUTs at
 future experiments. 
We also discuss five special scenarios with interesting possibilities.
With our generic formulae, we present all the GUT-scale 
gauge coupling relations and all the gaugino mass relations 
in the $SU(5)$ and $SO(10)$ models, and calculate the 
corresponding indices $k$. Especially, the index $k$ is $5/3$ 
in the traditional $SU(5)$ and $SO(10)$ models that 
have been studied extensively so far.
Furthermore, we discuss the field theory realization of the $U(1)$ flux
effects on the SM gauge kinetic functions in F-theory GUTs, and calculate
their indices $k$ as well.

\end{abstract}


\preprint{ACT-02-10, MIFP-10-07}

\maketitle

\section{Introduction}

%{\bf Introduction~--}~

Supersymmetry provides a natural solution to
the gauge hierarchy problem in the Standard Model (SM). In 
supersymmetric SMs with $R$ parity under which the SM particles
are even while their supersymmetric partners are odd, 
 the  gauge couplings for $SU(3)_C$, $SU(2)_L$ and $U(1)_Y$
gauge symmetries are unified at about $2\times 10^{16}$~GeV~\cite{Ellis:1990zq},
the lightest supersymmetric particle (LSP) like the neutralino can
be the cold dark matter candidate~\cite{Ellis:1983wd, Goldberg:1983nd}, 
and the electroweak precision constraints
can be evaded, etc. In particular, gauge coupling unification~\cite{Ellis:1990zq} 
strongly suggests Grand Unified Theories (GUTs), which  explain the
quantum numbers of the SM fermions and charge quantization. 
Thus, the great challenge is how to test the supersymmetric 
GUTs at the Large Hadron Collider (LHC), the future International 
Linear Collider (ILC), and other experiments.

In the supersymmetric SMs, supersymmetry is broken in
the hidden sector, and then its breaking effects
are mediated to the SM observable sector. However,
the relations between the supersymmetric particle
(sparticle) spectra and
the fundamental theories can be very complicated and model
dependent. Interestingly, comparing to the  supersymmetry 
breaking soft masses for squarks and sleptons, the gaugino masses  
have the simplest form and appear to be the least model 
dependent. With gravity mediated supersymmetry breaking
in GUTs, we have a universal gaugino mass $M_{1/2}$ at the 
GUT scale, which is called the minimal supergravity (mSUGRA)
scenario~\cite{mSUGRA}. Thus, we have the gauge coupling
relation and the gaugino mass
relation at the GUT scale $M_{\rm GUT}$:
\begin{eqnarray}
{{1}\over {\alpha_3}} ~=~  {{1}\over {\alpha_2}} 
~=~  {{1}\over {\alpha_1}} ~,~\,
\label{mSUGRA-C}
\end{eqnarray}
\begin{eqnarray}
{{M_3}\over {\alpha_3}} ~=~  {{M_2}\over {\alpha_2}} 
~=~  {{M_1}\over {\alpha_1}} ~,~\,
\label{mSUGRA}
\end{eqnarray}
where $\alpha_3$, $\alpha_2$, and $\alpha_1\equiv 5\alpha_{Y}/3$ are gauge
couplings respectively for $SU(3)_C$, $SU(2)_L$, 
and $U(1)_Y$ gauge symmetries, and 
$M_3$, $M_2$, and $M_1$ are  the  masses 
 for $SU(3)_C$, $SU(2)_L$, and $U(1)_Y$ 
gauginos, respectively. Interestingly, 
$1/\alpha_i$ and $M_i/\alpha_i$ satisfy the same 
equation $x_3=x_2=x_1$ at the GUT scale, which will be 
proved as a general result.
Because $M_i/\alpha_i$ are 
constant under one-loop renormalization group
equation (RGE) running, we obtain that
the above  gaugino mass relation in Eq.~(\ref{mSUGRA}) is valid
 from the GUT scale to the electroweak scale at one loop.
Note that the two-loop RGE running effects on gaugino masses
are very small, thus, we can test this gaugino mass relation
at the LHC and ILC where the gaugino masses can be 
measured~\cite{Cho:2007qv, Barger:1999tn}.
However,  the SM gauge couplings in GUTs need not
be unified at the GUT scale after the GUT gauge symmetry
breaking since the high-dimensional 
operators will contribute to the different gauge kinetic terms for
the $SU(3)_C$, $SU(2)_L$, and $U(1)_Y$ gauge 
symmetries~\cite{Hill:1983xh, Shafi:1983gz, Ellis:1984bm, Ellis:1985jn, Drees:1985bx}. 
Furthermore, we will have non-universal
gaugino masses at the GUT scale as 
well~\cite{Ellis:1985jn, Drees:1985bx, Anderson:1999uia,
Chamoun:2001in, Chakrabortty:2008zk, Martin:2009ad, Bhattacharya:2009wv, 
Feldman:2009zc, Chamoun:2009nd}.  In particular,
in the GUTs with large number of fields, the renormalization
effects significantly decrease the scale at which quantum
gravity becomes strong,  so, these high-dimensional operators
are indeed important and need to be considered 
seriously~\cite{Calmet:2008df}. 
Therefore, the key question is whether we still have the
gaugino mass relations that can be tested at
the LHC and ILC. It is amusing to  notice that the first systematic
studies for $SU(5)$ models
in the framework of $N=1$ supergravity for non-universal
gauge couplings and gaugino masses at the GUT scale were done
twenty-five years ago~\cite{Ellis:1985jn}.

On the other hand, in F-theory model building~\cite{Vafa:1996xn,
Donagi:2008ca, Beasley:2008dc, Beasley:2008kw, Donagi:2008kj,
Font:2008id, Jiang:2009zza, Blumenhagen:2008aw, Jiang:2009za,
Li:2009cy, Leontaris:2009wi, Li:2010mr}, the
GUT gauge fields are on the observable seven-branes which
wrap a del Pezzo $n$ surface $dP_n$ for the extra four space
dimensions. The
SM fermions and Higgs fields are on the complex codimension-one
curves (two-dimensional real subspaces) in  $dP_n$, and
the SM fermion Yukawa couplings arise from the intersections
of the SM fermion and Higgs curves. 
A brand new feature is that the $SU(5)$ gauge symmetry
can be broken down to the SM gauge symmetry
by turning on $U(1)_Y$ 
flux~\cite{Beasley:2008dc, Beasley:2008kw, Li:2009cy}, 
and the $SO(10)$  gauge 
symmetry can be broken down to the $SU(5)\times U(1)_X$
and $SU(3)\times SU(2)_L\times SU(2)_R\times U(1)_{B-L}$
gauge symmetries by turning on the $U(1)_X$ and $U(1)_{B-L}$
fluxes, respectively~\cite{Beasley:2008dc, Beasley:2008kw, 
Jiang:2009zza, Jiang:2009za, Font:2008id, Li:2009cy}. 
It has been shown that the gauge kinetic functions receive
the corrections from $U(1)$ fluxes~\cite{Donagi:2008kj, 
Blumenhagen:2008aw, Jiang:2009za, Li:2009cy}. Thus, whether we
can test F-theory GUT at the LHC and ILC is another
interesting question~\cite{Li:2010mr}.

In this paper, we consider the generalization of the mSUGRA (GmSUGRA).
In GUTs with gravity mediated supersymmetry
breaking,  we study for the first time the generic gauge 
coupling relations at the GUT scale, and the general 
gaugino mass relations which are valid
from the GUT scale to the electroweak scale at one loop. 
Interestingly, the gauge coupling relations and the
gaugino mass relations at the GUT scale are given by the same
equation. In other words,
$1/\alpha_i $ and $M_i/\alpha_i$ satisfy the same
equation at the GUT scale respectively for 
 the gauge coupling relations and the
gaugino mass relations. Thus, we define 
the index $k$ for these relations, which can be calculated in 
 GUTs and can be determined at the LHC and ILC. Therefore,
we present a concrete definition of the GUT scale in these theories, 
and suggest a new way to test general GUTs at the LHC and ILC.
Also, we discuss five special scenarios with interesting possibilities.
 With our generic formulae,
we present all the GUT-scale gauge coupling relations and all
 the gaugino mass relations in the $SU(5)$ and
$SO(10)$ models, and calculate the corresponding indices.  
Especially, in the 
traditional $SU(5)$ and $SO(10)$ models that have been 
studied extensively thus far, the index $k$ is $5/3$, 
which was first pointed out for $SU(5)$ models in Ref.~\cite{Ellis:1985jn}.
Moreover, we give the field theory realization of the $U(1)$ flux
effects on the SM gauge kinetic functions in F-theory GUTs. We find
 that in the $SU(5)$ and $SO(10)$ models respectively
with $U(1)_Y$ and $U(1)_{B-L}$ fluxes, the index $k$ is $5/3$,
while in the $SO(10)$ models with $U(1)_X$ flux, the gauge coupling 
relation and the gaugino 
mass relation are the same as these in the mSUGRA. Furthermore,
in four-dimensional
GUTs, the GUT gauge symmetry breaking may also affect the supersymmetry
breaking scalar masses, trilinear soft terms as well as
the SM fermion Yukawa couplings, which
will be studied elsewhere~\cite{TLDN-P}.

%%%%%%%%%%%%%%%%%%%%%%%%%%%%%%%%%%%%%%%%%%%%%%%%%%%%%%%%%%

%%%%%%%%%%%%%%%%%%%%%%%%%%%%%%%%%%%%%%%%%%%%%%%%%%%%%%%%%%

\section{Gauge Coupling Relations and Gaugino Mass Relations}

%{\bf Gaugino Mass Relations~--}~

After the GUT gauge symmetry breaking,
 we can parametrize the gauge kinetic functions
$f_3$, $f_2$ and $f_1$ respectively
for $SU(3)_C$, $SU(2)_L$ and $U(1)_Y$ gauge symmetries
at the GUT scale as follows
\begin{eqnarray}
f_i &=& \sum_m a'_{m}\tau_m + \epsilon \left (\sum_{n} a_{in} S_n \right)~,~\,  
\end{eqnarray}
where the first term is the original GUT gauge kinetic function, 
and the second term arise from the GUT gauge symmetry
breaking.  $\epsilon$ is
a small paramter close to the ratio between the GUT Higgs vacuum 
expectation value (VEV) and the fundamental scale $M_*$.
$\tau_m$ and $S_n$ are the hidden sector fields whose $F$-terms
may break supersymmetry. In particular, for $a_{1n}=a_{2n}=a_{3n}$,
the gauge coupling relation at the GUT scale and the gaugino mass relation
are the same as these in the mSUGRA. 

{\bf Theorem.} If there exist three real numbers $b_i$ such that
$\sum_{i=1}^3 b_i f_i = 0$, we have the following gauge coupling relation
at the GUT scale 
\begin{eqnarray}
{{b_3}\over {\alpha_3}} + {{b_2}\over {\alpha_2}}
+ {{b_1}\over {\alpha_1}} =0~.~\,  
\label{GaugeCR}
\end{eqnarray}
Using one-loop RGE running, we have the following gaugino mass 
relation which is  renormalization scale invariant from
 the GUT scale to the electroweak scale at one loop
\begin{eqnarray}
{{b_3 M_3}\over {\alpha_3}} + {{b_2 M_2}\over {\alpha_2}}
+ {{b_1 M_1}\over {\alpha_1}} =0~.~\,  
\label{GauginoMR}
\end{eqnarray}

{\bf Proof.}  Because $f_i=1/(4\pi\alpha_i)$, the
gauge coupling relation in Eq.~(\ref{GaugeCR}) at the GUT scale
is obtained automatically. 

From $\sum_{i=1}^3 b_i f_i = 0$, we have
\begin{eqnarray}
\sum_{i=1}^3 b_i ~=~0~,~~~\sum_{i=1}^3 b_i a_{in} ~=~0~.~\,
\label{Conditions}  
\end{eqnarray}

Assuming that the F-terms of $\tau_m$ and $S_n$ break supersymmetry,
we obtain the ratios between the gaugino masses and gauge couplings
\begin{eqnarray}
{{M_i}\over {\alpha_i}}~=~ 4\pi \left[\sum_m  a'_m F^{\tau_m}
+\epsilon \left(\sum_n  a_{in} F^{S_n}\right) \right]~.~\,
\end{eqnarray}
Using Eq.~(\ref{Conditions}), we obtain the gaugino mass relation
given in Eq.~(\ref{GauginoMR}) at the GUT scale. Because 
$M_i/\alpha_i$ are invariant under one-loop RGE running, 
we prove the theorem. The gaugino mass relation will have very 
small deviation due to the two-loop RGE running~\cite{Li:2010mr}.

Interestingly, the GUT-scale gauge coupling relation in Eq.~(\ref{GaugeCR}) 
and the gaugino mass
relation in  Eq.~(\ref{GauginoMR})  give the same equation as follows
\begin{eqnarray}
b_3 x_3 + b_2 x_2 + b_1 x_1 ~=~0~.~\,  
\end{eqnarray}
In other words, $1/\alpha_i$ and $M_i/\alpha_i$ at the GUT scale
satisfy the same equation respectively for the gauge coupling relation
and the gaugino mass relation.
Thus, we can  define 
the GUT scale in these theories:

{\bf Definition.} The GUT scale is the scale at which 
$1/\alpha_i$ and $M_i/\alpha_i$
satisfy the same equation respectively for the gauge coupling
relation and the gaugino mass relation.

For simplicity, we consider two supersymmetry breaking
fields $\tau$ and $S$. The generic gauge kinetic 
function can be parametrized as follows
\begin{eqnarray}
f_i &=& \tau + \epsilon a_i  S~.~\,  
\label{GKF-1}
\end{eqnarray}
If $a_1=a_2=a_3$, similar to the mSUGRA, we obtain the  
GUT-scale gauge coupling
relation in  Eq.~(\ref{mSUGRA-C}) and the 
gaugino mass relation
in Eq.~(\ref{mSUGRA}).

If there exists at least one $a_i \not= a_j $ for $i\not= j$,
  we obtain the generic solution for $b_i$ up to a scale factor
\begin{eqnarray}
b_1 ~=~ a_2-a_3~,~~ b_2~=~ a_3-a_1~,~~b_3~=~ a_1-a_2~.~\,  
\end{eqnarray}
Using our theorem, we obtain the gauge coupling
relation at the GUT scale
\begin{eqnarray}
{{a_1-a_2}\over {\alpha_3}} + {{a_3-a_1}\over {\alpha_2}}
+ {{a_2-a_3}\over {\alpha_1}} =0~.~\,  
\label{CouplingR}
\end{eqnarray}
In addition, we obtain the gaugino mass relation
which is valid from the GUT scale to the electroweak scale
under one-loop RGE running
\begin{eqnarray}
{{(a_1-a_2) M_3}\over {\alpha_3}} + {{(a_3-a_1) M_2}\over {\alpha_2}}
+ {{(a_2-a_3) M_1}\over {\alpha_1}} =0~.~\,  
\label{MassR}
\end{eqnarray}

Except the mSUGRA, we always have $a_1 \not= a_2$ or $a_1\not=a_3$
in GmSUGRA in the following discussions. Thus, we can rewrite the
GUT-scale gauge coupling relation and 
the gaugino 
mass relation as follows
\begin{eqnarray}
{{1}\over {\alpha_2}} - {{1}\over {\alpha_3}} 
~=~k \left( {{1}\over {\alpha_1}} 
- {{1}\over {\alpha_3}} \right) ~,~\,
\label{GCRelation}
\end{eqnarray}
\begin{eqnarray}
{{M_2}\over {\alpha_2}} - {{M_3}\over {\alpha_3}} 
~=~k \left( {{M_1}\over {\alpha_1}} 
- {{M_3}\over {\alpha_3}} \right) ~,~\,
\label{GMRelation}
\end{eqnarray}
where $k$ is the index of these relations, and is defined as follows
\begin{eqnarray}
k ~\equiv~ {{a_2-a_3}\over {a_1-a_3}}~.~\,
\label{index}
\end{eqnarray}
Because $M_i/\alpha_i$ are renormalization scale invariant under one-loop RGE running
and can be calculated from the LHC and ILC experiments,
$k$  can be determined at the low energy as well. Therefore, we can test GUTs since
its $k$ can be calculated. 
Although $k$ is not well defined in the
mSUGRA, we symbolically define the index $k$ for mSUGRA as $k=0/0$.
In other words, for $k=0/0$, we have the gauge coupling relation
 at the GUT scale given by Eq.~(\ref{mSUGRA-C}), 
and the gaugino mass relation given by Eq.~(\ref{mSUGRA}).
In addition, the concrete GUT scale can be redefined as follows:

{\bf Definition.} The GUT scale is the scale at which the gauge coupling
relation and the gaugino mass relation have the same index  $k$.

Because the GUT gauge couplings should be positive and finite, we
obtain that ${\rm Re} \tau  > 0$. Let us consider five
 special cases in the following: \\

(1)  ${\rm Re} S\not=0$, $F^{\tau}\not=0$, $F^S=0$. \\

In this case, the gauge coupling relation at the GUT scale
 is still given by Eq.~(\ref{CouplingR})
or Eq.~(\ref{GCRelation}). However, the gaugino mass relation is given
by the mSUGRA gaugino mass relation in Eq.~(\ref{mSUGRA}).
This implies that even if we obtain the mSUGRA gaugino mass relation
at the LHC and ILC, we may still have the non-unified SM gauge couplings 
at the GUT scale. Unfortunately, we can not calculate $k$ in this case
 at the LHC and ILC.  \\

(2)  ${\rm Re} S \not=0$, $F^{\tau}=0$, $F^S\not=0$. \\

This case has been studied carefully 
in Refs.~\cite{Anderson:1999uia, Chamoun:2001in, Chakrabortty:2008zk, 
Martin:2009ad, Bhattacharya:2009wv, Feldman:2009zc, Chamoun:2009nd}. In this case, 
the gauge coupling relation at the GUT scale 
is still given by Eq.~(\ref{CouplingR})
or Eq.~(\ref{GCRelation}), and the gaugino mass relation is 
given by Eq.~(\ref{MassR}) or Eq.~(\ref{GMRelation}). In particular,
  for $a_i\not= 0$ we obtain the gaugino mass relation
\begin{eqnarray}
{{M_3}\over {a_3\alpha_3}} ~=~  {{M_2}\over {a_2\alpha_2}} 
~=~  {{M_1}\over {a_1\alpha_1}} ~.~\, 
\label{Ftm=0}
\\ \nonumber
\end{eqnarray}

(3)  ${\rm Re} S=0$, $F^{\tau}\not=0$, $F^S\not=0$. \\

In this case, the gauge coupling relation at the GUT scale is given by 
the mSUGRA gauge coupling relation in Eq.~(\ref{mSUGRA-C}), while
the gaugino mass relation is given by Eq.~(\ref{MassR})
or Eq.~(\ref{GMRelation}). Thus, 
 even if we obtain the non-universal gaugino mass relation
from the LHC and ILC, we may still have the gauge coupling unification
at the GUT scale. \\

(4)  ${\rm Re} S=0$, $F^{\tau}\not=0$, $F^S=0$. \\

This case is the same as the mSUGRA. \\

(5)  ${\rm Re} S=0$, $F^{\tau}=0$, $F^S\not=0$. \\

In this case,  the gauge coupling relation at the GUT scale is given 
by Eq.~(\ref{mSUGRA-C}),
while the gaugino mass relation is 
given by Eq.~(\ref{MassR}) or Eq.~(\ref{GMRelation}). 
Also, the  gaugino mass relation for $a_i\not= 0$ is given
by Eq.~(\ref{Ftm=0}) as well.

\section{Grand Unified Theories}

%{\bf Grand Unified Theories~--}~

In four-dimensional GUTs, the non-universal SM
gauge kinetic function can be generated after GUT gauge
symmetry breaking by the high-dimensional 
operators~\cite{Hill:1983xh, Shafi:1983gz, Ellis:1984bm, Ellis:1985jn, 
Drees:1985bx, Anderson:1999uia, Chamoun:2001in, Chakrabortty:2008zk, 
Martin:2009ad, Bhattacharya:2009wv, Feldman:2009zc, Chamoun:2009nd}.
The generic gauge kinetic function in the superpotential is 
\begin{eqnarray}
W \supset {1\over 2}  {\rm Tr} \left[ W^a W^b 
\left(\tau \delta_{ab} + \lambda{{\Phi_{ab}}\over {M_*}}S\right) \right]~,~\,
\end{eqnarray}
where $\lambda$ is the Yukawa coupling constant, and
$\Phi_{ab}$ transforms as the symmetric product of 
two adjoint representations. After $\Phi_{ab}$ obtains
a VEV, we obtain the gauge kinetic functions in
Eq.~(\ref{GKF-1}) where $\epsilon$ is the product of $\lambda$,
the VEV of $\Phi_{ab}$, and suitable normalization factors.

First, let us study the $SU(5)$ models. The symmetric
product of the adjoint representation ${\mathbf{24}}$ 
of $SU(5)$ can be decomposed into irreducible representations
of $SU(5)$ as follows
\begin{eqnarray}
({\mathbf{24}} \times {\mathbf{24}})_{\rm symmetric}
&=& {\mathbf{1}} \oplus  {\mathbf{24}} \oplus {\mathbf{75}}
\oplus  {\mathbf{200}}~.~\,
\end{eqnarray}
We present $a_i$ and index $k$ for each irreducible
representation  in Table~\ref{SU(5)-I}.
Thus, using our general formulae in Section II, we have all the GUT-scale
gauge coupling relations and all the gaugino mass relations
in $SU(5)$ models. Especially, 
in the traditional $SU(5)$ models that have been studied extensively
so far, the GUT Higgs field is in the representation ${\mathbf{24}}$,
and then the index $k$ is $5/3$.
By the way, the gaugino mass relations for the Higgs fields
in the representations ${\mathbf{24}}$ and ${\mathbf{75}}$ have
been studied previously~\cite{Ellis:1985jn}.

%%%%%%%%%%%%%%%%%%%%%%%%%%%%%%%%%%%%%%%%%%%%%%%%%%%%%%%%%%%%%%%%%%%%%

%%%%%%%%%%%%%%%%%%%%%%%%%%%%%%%%%%%%%%%%%%%%%%%%%%%%%%%%%%%%%%%%%%%%%

\begin{table}[htb]
\begin{center}
\begin{tabular}{|c|c|c|c|c|}
\hline
$SU(5)$  & $a_1$ & $a_2$ & 
 $a_3$   &  $k$ \\
\hline
${\mathbf{1}}$ & $1$ & $1$ & $1$ & $0/0$ \\
\hline
${\mathbf{24}}$ & $-1/2$ & $-3/2$ & $1$ & $5/3$ \\
\hline
${\mathbf{75}}$ & $-5$ & $3$ & $1$ & $-1/3$ \\
\hline
${\mathbf{200}}$ & $10$ & $2$ & $1$ & $1/9$ \\
\hline
\end{tabular}
\end{center}
\caption{$a_i$ and $k$ for each irreducible
representation in $SU(5)$ models.}
\label{SU(5)-I}
\end{table}

%%%%%%%%%%%%%%%%%%%%%%%%%%%%%%%%%%%%%%%%%%%%%%%%%%%%%%%%%%%%%%%%%%%%%

%%%%%%%%%%%%%%%%%%%%%%%%%%%%%%%%%%%%%%%%%%%%%%%%%%%%%%%%%%%%%%%%%%%%%

Second, let us consider the $SO(10)$ models. The symmetric
product of the adjoint representation ${\mathbf{45}}$ 
of $SO(10)$ can be decomposed into irreducible representations
of $SO(10)$ as follows
\begin{eqnarray}
({\mathbf{45}} \times {\mathbf{45}})_{\rm symmetric}
&=& {\mathbf{1}} \oplus  {\mathbf{54}} \oplus {\mathbf{210}}
\oplus  {\mathbf{770}}~.~\,
\end{eqnarray}
The $SO(10)$ models can be broken down to the Georgi-Glashow 
$SU(5)\times U(1)$ models, the flipped $SU(5)\times U(1)_X$ models,
and the Pati-Salam $SU(4)_C\times SU(2)_L\times SU(2)_R$ models. 
We present $a_i$ and indices $k$ for each irreducible
representation in Table~\ref{SO(10)-GG}, Table~\ref{SO(10)-F}
 and Table~\ref{SO(10)-PS}  for the $SO(10)$ models
whose gauge symmetries are broken down to the Georgi-Glashow 
$SU(5)\times U(1)$ gauge symmetries, the flipped $SU(5)\times U(1)_X$ 
gauge symmetries, and
the Pati-Salam $SU(4)_C\times SU(2)_L\times SU(2)_R$ gauge symmetries,
respectively. We emphasize that our numbers $a_i$ in 
 Table~\ref{SO(10)-GG}, Table~\ref{SO(10)-F} and Table~\ref{SO(10)-PS}
are the same as the results obtained in the corresponding 
Tables in Ref.~\cite{Martin:2009ad}.
Thus, using our generical formulae in Section II, 
we have all the GUT-scale
gauge coupling relations and all the gaugino mass relations
in $SO(10)$ models.

In the traditional $SO(10)$ models that have been studied extensively
so far, the GUT Higgs fields are in the representations ${\mathbf{45}}$
as well as $\mathbf{16}$ and  $\mathbf{\overline{16}}$~\cite{Georgi:1979dq}. 
Thus, the above
discussions can not be applied directly. In this case, the discussions
on the GUT-scale gauge coupling relation and  
the gaugino mass relation are similar to these in the field theory
realization of the F-theory $SO(10)$ models with $U(1)_{B-L}$ flux.
As discussed in the following, the index $k$ in the traditional
 $SO(10)$ models is $5/3$ as well.

%%%%%%%%%%%%%%%%%%%%%%%%%%%%%%%%%%%%%%%%%%%%%%%%%%%%%%%%%%%%%%%%%%%%%

%%%%%%%%%%%%%%%%%%%%%%%%%%%%%%%%%%%%%%%%%%%%%%%%%%%%%%%%%%%%%%%%%%%%%

\begin{table}[htb]
\begin{center}
\begin{tabular}{|c|c|c|c|c|c|}
\hline
$SO(10)$ & $SU(5)\times U(1)$ & $a_1$ & $a_2$ & 
 $a_3$   &  $k$ \\
\hline
${\mathbf{1}}$ & $(\mathbf{1}, \mathbf{0})$ & $1$ & $1$ & $1$ & $0/0$ \\
\hline
${\mathbf{54}}$ & $(\mathbf{24}, \mathbf{0})$ &   $-1/2$ & $-3/2$ & $1$ & $5/3$ \\
\hline
 & $(\mathbf{1}, \mathbf{0})$ & $1$ & $1$ & $1$ & $0/0$ \\
${\mathbf{210}}$ &  $(\mathbf{24}, \mathbf{0})$ &   $-1/2$ & $-3/2$ & $1$ & $5/3$ \\
 &  $(\mathbf{75}, \mathbf{0})$ &  $-5$ & $3$ & $1$ & $-1/3$ \\
\hline
 & $(\mathbf{1}, \mathbf{0})$ & $1$ & $1$ & $1$ & $0/0$ \\
${\mathbf{770}}$ &  $(\mathbf{24}, \mathbf{0})$ &   $-1/2$ & $-3/2$ & $1$ & $5/3$ \\
 &  $(\mathbf{75}, \mathbf{0})$ &  $-5$ & $3$ & $1$ & $-1/3$ \\
 &  $(\mathbf{200}, \mathbf{0})$ &    $10$ & $2$ & $1$ & $1/9$ \\
\hline
\end{tabular}
\end{center}
\caption{$a_i$ and $k$ for each irreducible
representation in $SO(10)$ models whose gauge symmetry
is broken down to the Georgi-Glashow $SU(5)\times U(1)$
gauge symmetries.}
\label{SO(10)-GG}
\end{table}

%%%%%%%%%%%%%%%%%%%%%%%%%%%%%%%%%%%%%%%%%%%%%%%%%%%%%%%%%%%%%%%%%%%%%

%%%%%%%%%%%%%%%%%%%%%%%%%%%%%%%%%%%%%%%%%%%%%%%%%%%%%%%%%%%%%%%%%%%%%

%%%%%%%%%%%%%%%%%%%%%%%%%%%%%%%%%%%%%%%%%%%%%%%%%%%%%%%%%%%%%%%%%%%%%

%%%%%%%%%%%%%%%%%%%%%%%%%%%%%%%%%%%%%%%%%%%%%%%%%%%%%%%%%%%%%%%%%%%%%

\begin{table}[htb]
\begin{center}
\begin{tabular}{|c|c|c|c|c|c|}
\hline
$SO(10)$ & $SU(5)\times U(1)_X$ & $a_1$ & $a_2$ & 
 $a_3$   &  $k$ \\
\hline
${\mathbf{1}}$ & $(\mathbf{1}, \mathbf{0})$ & $1$ & $1$ & $1$ & $0/0$ \\
\hline
${\mathbf{54}}$ & $(\mathbf{24}, \mathbf{0})$ &   $-1/2$ & $-3/2$ & $1$ & $5/3$ \\
\hline
 & $(\mathbf{1}, \mathbf{0})$ & $-19/5$ & $1$ & $1$ & $0$ \\
${\mathbf{210}}$ &  $(\mathbf{24}, \mathbf{0})$ &   $7/10$ & $-3/2$ & $1$ & $25/3$ \\
 &  $(\mathbf{75}, \mathbf{0})$ &  $-1/5$ & $3$ & $1$ & $-5/3$ \\
\hline
 & $(\mathbf{1}, \mathbf{0})$ & $77/5$ & $1$ & $1$ & $0$ \\
${\mathbf{770}}$ &  $(\mathbf{24}, \mathbf{0})$ &   $-101/10$ & $-3/2$ & $1$ & $25/111$ \\
 &  $(\mathbf{75}, \mathbf{0})$ &  $-1/5$ & $3$ & $1$ & $-5/3$ \\
 &  $(\mathbf{200}, \mathbf{0})$ &    $2/5$ & $2$ & $1$ & $-5/3$ \\
\hline
\end{tabular}
\end{center}
\caption{$a_i$ and $k$ for each irreducible
representation in $SO(10)$ models whose gauge symmetry
is broken down to the flipped $SU(5)\times U(1)_X$
gauge symmetries.}
\label{SO(10)-F}
\end{table}

%%%%%%%%%%%%%%%%%%%%%%%%%%%%%%%%%%%%%%%%%%%%%%%%%%%%%%%%%%%%%%%%%%%%%

%%%%%%%%%%%%%%%%%%%%%%%%%%%%%%%%%%%%%%%%%%%%%%%%%%%%%%%%%%%%%%%%%%%%%

%%%%%%%%%%%%%%%%%%%%%%%%%%%%%%%%%%%%%%%%%%%%%%%%%%%%%%%%%%%%%%%%%%%%%

%%%%%%%%%%%%%%%%%%%%%%%%%%%%%%%%%%%%%%%%%%%%%%%%%%%%%%%%%%%%%%%%%%%%%

\begin{table}[htb]
\begin{center}
\begin{tabular}{|c|c|c|c|c|c|}
\hline
$SO(10)$ & $SU(4)_C\times SU(2)_L \times SU(2)_R$ & $a_1$ & $a_2$ & 
 $a_3$   &  $k$ \\
\hline
${\mathbf{1}}$ & $(\mathbf{1}, \mathbf{1}, \mathbf{1})$ & $1$ & $1$ & $1$ & $0/0$ \\
\hline
${\mathbf{54}}$ & $(\mathbf{1}, \mathbf{1}, \mathbf{1})$  &   $-1/2$ & $-3/2$ & $1$ & $5/3$ \\
\hline
 & $(\mathbf{1}, \mathbf{1}, \mathbf{1})$   & $-3/5$ & $1$ & $0$ & $-5/3$ \\
${\mathbf{210}}$ &  $(\mathbf{15}, \mathbf{1}, \mathbf{1})$ &   $-4/5$ & $0$ & $1$ & $5/9$ \\
 & $(\mathbf{15}, \mathbf{1}, \mathbf{3})$  &  $1$ & $0$ & $0$ & $0$ \\
\hline
 & $(\mathbf{1}, \mathbf{1}, \mathbf{1})$ & $19/10$ & $5/2$ & $1$ & $5/3$ \\
${\mathbf{770}}$ & $(\mathbf{1}, \mathbf{1}, \mathbf{5})$  &   $1$ & $0$ & $0$ & $0$ \\
 & $(\mathbf{15}, \mathbf{1}, \mathbf{3})$  &  $1$ & $0$ & $0$ & $0$ \\
 & $(\mathbf{84}, \mathbf{1}, \mathbf{1})$  &    $32/5$ & $0$ & $1$ & $-5/27$ \\
\hline
\end{tabular}
\end{center}
\caption{$a_i$ and $k$ for each irreducible
representation in $SO(10)$ models whose gauge symmetry
is broken down to the Pati-Salam $SU(4)_C\times SU(2)_L \times SU(2)_R$
gauge symmetries.}
\label{SO(10)-PS}
\end{table}

%%%%%%%%%%%%%%%%%%%%%%%%%%%%%%%%%%%%%%%%%%%%%%%%%%%%%%%%%%%%%%%%%%%%%

%%%%%%%%%%%%%%%%%%%%%%%%%%%%%%%%%%%%%%%%%%%%%%%%%%%%%%%%%%%%%%%%%%%%%

%%%%%%%%%%%%%%%%%%%%%%%%%%%%%%%%%%%%%%%%%%%%

\section{F-Theory GUTs}

%{\bf F-Theory GUTs~--}~

 We consider the field theory realization 
of the $U(1)$ flux
effects on the SM gauge kinetic functions in F-theory GUTs, and study their
GUT-scale gauge coupling relations and their
gaugino mass relations~\cite{Donagi:2008kj, Blumenhagen:2008aw, 
Jiang:2009za, Li:2009cy}. In the F-theory $SU(5)$ models, 
the $SU(5)$ gauge symmetry
is broken down to the SM gauge symmetry by turning on
the $U(1)_Y$ flux. To realize the $U(1)_Y$
flux corrections to the SM gauge kinetic functions in
the four-dimensional $SU(5)$ models,
we consider the following superpotential term
for the $SU(5)$ gauge kinetic function
\begin{eqnarray}
W\supset {1\over 2}  {\rm Tr} \left[ W^a W^b 
\left(\tau \delta_{ab} + 
\left({Z^2\delta_{ab}+ \lambda{\Phi_{a} \Phi_{b}}\over {M^2_*}}\right) S
\right) \right],~\,
\end{eqnarray}
where $Z$ is a SM singlet Higgs field, and $\Phi_a$ and $\Phi_b$ 
are the Higgs
fields in the adjoint representation of $SU(5)$ which have
VEVs along the $U(1)_Y$ direction. 
Five stacks of seven-branes give us $U(5)$ symmetry,
thus, the $Z^2$ term is similar to the flux for the
global $U(1)$ of $U(5)$,
and the $\Phi_{a} \Phi_{b}$ term is similar to the
$U(1)_Y$ flux~\cite{Donagi:2008kj, Blumenhagen:2008aw}.
After $SU(5)$ gauge symmetry is broken down to the SM
gauge symmetry, with suitable definition of $\epsilon$, 
 we obtain~\cite{Donagi:2008kj, Blumenhagen:2008aw}
\begin{eqnarray}
a_1 = {1\over 2}\left(\alpha+{6\over 5}\right)~,~
a_2={1\over 2}\left( \alpha + 2 \right)~,~ a_3= {1\over 2}
\alpha ~,~\,
\end{eqnarray}
where $\alpha$ is a real number. In F-theory models,
$\alpha$ should be quantized due to flux quantization.
Thus, using Eqs.~(\ref{CouplingR}) and (\ref{MassR})
or Eqs.~(\ref{GCRelation}) and (\ref{GMRelation}),
we can easily obtain the gauge coupling relation
at the GUT scale and the gaugino mass relation whose
index $k$ is $5/3$~\cite{Li:2010mr}.

In the F-theory $SO(10)$ models where the $SO(10)$ gauge symmetry
is broken down to the flipped $SU(5)\times U(1)_X$ gauge symmetry
by turning on the $U(1)_X$ flux~\cite{Beasley:2008dc, Beasley:2008kw,
Jiang:2009zza, Jiang:2009za}, we can show that the gauge kinetic
functions for $SU(5)$ and $U(1)_X$ are exactly the 
same at the unification scale~\cite{Jiang:2009za}. 
In the field theory realization, we consider
the following superpotential term for
the $SO(10)$ gauge kinetic function
\begin{eqnarray}
W\supset {1\over 2}  {\rm Tr} \left[ W^a W^b 
\left(\tau \delta_{ab} + 
\lambda  {{\Phi_{a} \Phi_{b}}\over {M^2_*}} S
\right) \right]~,~\,
\label{SO(10)-GKF}
\end{eqnarray}
where $\Phi_a$ and $\Phi_b$ are the Higgs fields 
in the adjoint representation of $SO(10)$. To have similar effects
as the $U(1)_X$ flux, $\Phi_a$ and $\Phi_b$ obtain VEVs along the 
$U(1)_X$ direction. Thus, with suitable definition of $\epsilon$,
we get~\cite{Jiang:2009za}
\begin{eqnarray}
a_1~=~a_2~=~a_3~=~1 ~.~\,
\end{eqnarray}
Therefore,
similar to the mSUGRA, we obtain the gauge coupling unification at the GUT scale
in Eq.~(\ref{mSUGRA-C}) and the gaugino mass relation
in Eq.~(\ref{mSUGRA}), {\it i.e.}, we have $k=\infty$.

In the  F-theory $SO(10)$ models, the $SO(10)$ gauge symmetry can also 
be broken down to the 
$SU(3)_C\times SU(2)_L\times SU(2)_R\times U(1)_{B-L}$
gauge symmetry by turning on the $U(1)_{B-L}$ 
flux~\cite{Font:2008id, Li:2009cy}. To realize the $U(1)_{B-L}$ 
flux corrections
to the SM gauge kinetic functions in four-dimensional $SO(10)$ models, 
we still consider the  superpotential term in Eq.~(\ref{SO(10)-GKF}),
where $\Phi_a$ and $\Phi_b$ obtain VEVs along the 
$U(1)_{B-L}$ direction. Thus, with suitable definition of $\epsilon$,
we get~\cite{Li:2009cy} 
\begin{eqnarray}
a_1~=~{2\over 5}~,~~
a_2~=~0~,~~ a_3~=~1 ~.~\,
\end{eqnarray}
Therefore, using Eqs.~(\ref{CouplingR}) and (\ref{MassR})
or Eqs.~(\ref{GCRelation}) and (\ref{GMRelation}),
we can easily obtain the gauge coupling relation
at the GUT scale and the gaugino mass relation whose
index $k$ is $5/3$~\cite{Li:2010mr}.

%%%%%%%%%%%%%%%%%%%%%%%%%%%%%%%%%%%%%%%%%%%%%%%%%%%%%%%%%%

%%%%%%%%%%%%%%%%%%%%%%%%%%%%%%%%%%%%%%%%%%%%%%%%%%%%%%%%%%

%%%%%%%%%%%%%%%%%%%%%%%%%%%%%%%%%%%%%%%%%%%%%%%%%%%%%%%%%%

%%%%%%%%%%%%%%%%%%%%%%%%%%%%%%%%%%%%%%%%%%%%%%%%%%%%%%%%%%

\section{Conclusions}

%{\bf Conclusion~--}~

In  GUTs with gravity mediated 
supersymmetry breaking,
we considered the generic gauge coupling relations at
the GUT scale, and  the general gaugino mass relations 
which are valid from the GUT scale to the electroweak scale 
at one loop. 
Interestingly, the gauge coupling relations and the
gaugino mass relations at the  GUT-scale are given by the same
equation, {\it i.e.},
$1/\alpha_i $ and $M_i/\alpha_i$ satisfy the same
equation respectively for the gauge coupling relations
and the gaugino mass relations. Thus, we define 
the index $k$ for these relations.
Because  the index $k$  can be calculated in GUTs and 
can be determined at the LHC and future ILC, 
we gave a concrete definition of the GUT scale in these theories, 
and suggested a new way to test general GUTs at the future experiments.
We also dicussed five special scenarios with interesting possibilities.
With our generic formulae, we presented all the GUT-scale gauge coupling 
relations and all the gaugino mass relations in the $SU(5)$ and
$SO(10)$ models, and calculated the corresponding indices. 
In particular, the index $k$ is $5/3$~\cite{Ellis:1985jn} in the 
traditional $SU(5)$ and $SO(10)$ models that have been 
studied extensively so far.
Moreover, we studied the field theory realization of the $U(1)$ flux
effects on the SM gauge kinetic functions in F-theory GUTs. We found
 that in the $SU(5)$ and $SO(10)$ models respectively
with $U(1)_Y$ and $U(1)_{B-L}$ fluxes, the index $k$ is $5/3$,
while in the $SO(10)$ models with $U(1)_X$ flux,
the GUT-scale gauge coupling relation and gaugino mass relation
are the same as these in  mSUGRA.
In short, the gaugino mass relation with index $k=5/3$~\cite{Ellis:1985jn}
 definitely deserve further detail study.

\begin{acknowledgments}

%{\bf Acknowledgments~--}~

This research was supported in part 
by  the DOE grant DE-FG03-95-Er-40917 (TL and DVN),
by the Natural Science Foundation of China 
under grant No. 10821504 (TL),
and by the Mitchell-Heep Chair in High Energy Physics (TL).

\end{acknowledgments}

%%%%%%%%%%%%%%%%%%%%%%%%%%%%%%%%%%%%%%%%%%%%%%%%%%%%%%%%%%%%%%%%%%%%%%%%%%%%


\begin{thebibliography}{99}



%%%%%%%%%%%%%%%%%%%%%%%%%%%%%%%%%%%%%%%%%%%%%%%%%%%%%%%%%%%%%%%%%%%%%%

%%%%%%%%%%%%%%%%%%%%%%%%%%%%%%%%%%%%%%%%%%%%%%%%%%%%%%%%%%%%%%%%%%%%%%




%\cite{Ellis:1990zq}
\bibitem{Ellis:1990zq}
  J.~R.~Ellis, S.~Kelley and D.~V.~Nanopoulos,
  %``Precision Lep Data, Supersymmetric Guts And String Unification,''
  Phys.\ Lett.\  B {\bf 249}, 441 (1990);
  %%CITATION = PHLTA,B249,441;%%
%``Probing the desert using gauge coupling unification,''
  Phys.\ Lett.\  B {\bf 260}, 131 (1991);
  %%CITATION = PHLTA,B260,131;%%
 U.~Amaldi, W.~de Boer and H.~Furstenau,
  %``Comparison of grand unified theories with electroweak and strong coupling
  %constants measured at LEP,''
  Phys.\ Lett.\  B {\bf 260}, 447 (1991);
  %%CITATION = PHLTA,B260,447;%%
 P.~Langacker and M.~X.~Luo,
  %``Implications of precision electroweak experiments for $M_t$, $\rho_{0}$,
  %$\sin^2\theta_W$ and grand unification,''
  Phys.\ Rev.\  D {\bf 44}, 817 (1991).
  %%CITATION = PHRVA,D44,817;%%


%\cite{Ellis:1983wd}
\bibitem{Ellis:1983wd}
  J.~R.~Ellis, J.~S.~Hagelin, D.~V.~Nanopoulos and M.~Srednicki,
  %``Search For Supersymmetry At The Anti-P P Collider,''
  Phys.\ Lett.\  B {\bf 127}, 233 (1983);
  %%CITATION = PHLTA,B127,233;%%
J.~R.~Ellis, J.~S.~Hagelin, D.~V.~Nanopoulos, K.~A.~Olive and M.~Srednicki,
  %``Supersymmetric relics from the big bang,''
  Nucl.\ Phys.\  B {\bf 238}, 453 (1984).
  %%CITATION = NUPHA,B238,453;%%


%\cite{Goldberg:1983nd}
\bibitem{Goldberg:1983nd}
  H.~Goldberg,
  %``Constraint on the photino mass from cosmology,''
  Phys.\ Rev.\ Lett.\  {\bf 50}, 1419 (1983)
  [Erratum-ibid.\  {\bf 103}, 099905 (2009)].
  %%CITATION = PRLTA,50,1419;%%



%\cite{mSUGRA}
\bibitem{mSUGRA}
  A.~H.~Chamseddine, R.~L.~Arnowitt and P.~Nath,
  %``Locally Supersymmetric Grand Unification,''
  Phys.\ Rev.\ Lett.\  {\bf 49}, 970 (1982);
  %%CITATION = PRLTA,49,970;%%
H.~P.~Nilles,
  %``Dynamically Broken Supergravity And The Hierarchy Problem,''
  Phys.\ Lett.\  B {\bf 115}, 193 (1982);
  %%CITATION = PHLTA,B115,193;%%
L.~E.~Ibanez,
  %``Locally Supersymmetric SU(5) Grand Unification,''
  Phys.\ Lett.\  B {\bf 118}, 73 (1982);
  %%CITATION = PHLTA,B118,73;%%
R.~Barbieri, S.~Ferrara and C.~A.~Savoy,
  %``Gauge Models With Spontaneously Broken Local Supersymmetry,''
  Phys.\ Lett.\  B {\bf 119}, 343 (1982);
  %%CITATION = PHLTA,B119,343;%%
H.~P.~Nilles, M.~Srednicki and D.~Wyler,
  %``Weak Interaction Breakdown Induced By Supergravity,''
  Phys.\ Lett.\  B {\bf 120}, 346 (1983);
  %%CITATION = PHLTA,B120,346;%%
J.~R.~Ellis, D.~V.~Nanopoulos and K.~Tamvakis,
  %``Grand Unification In Simple Supergravity,''
  Phys.\ Lett.\  B {\bf 121}, 123 (1983);
  %%CITATION = PHLTA,B121,123;%%
J.~R.~Ellis, J.~S.~Hagelin, D.~V.~Nanopoulos and K.~Tamvakis,
  %``Weak Symmetry Breaking By Radiative Corrections In Broken Supergravity,''
  Phys.\ Lett.\  B {\bf 125}, 275 (1983);
  %%CITATION = PHLTA,B125,275;%%
L.~J.~Hall, J.~D.~Lykken and S.~Weinberg,
  %``Supergravity As The Messenger Of Supersymmetry Breaking,''
  Phys.\ Rev.\  D {\bf 27}, 2359 (1983).
  %%CITATION = PHRVA,D27,2359;%%



%\cite{Cho:2007qv}
\bibitem{Cho:2007qv}
  W.~S.~Cho, K.~Choi, Y.~G.~Kim and C.~B.~Park,
  %``Gluino Stransverse Mass,''
  Phys.\ Rev.\ Lett.\  {\bf 100}, 171801 (2008);
%  [arXiv:0709.0288 [hep-ph]].
  %%CITATION = PRLTA,100,171801;%%
%\cite{Nojiri:2008hy}
%\bibitem{Nojiri:2008hy}
  M.~M.~Nojiri, Y.~Shimizu, S.~Okada and K.~Kawagoe,
  %``Inclusive transverse mass analysis for squark and gluino mass
  %determination,''
  JHEP {\bf 0806}, 035 (2008).
%  [arXiv:0802.2412 [hep-ph]].
  %%CITATION = JHEPA,0806,035;%%


%\cite{Barger:1999tn}
\bibitem{Barger:1999tn}
  V.~D.~Barger, T.~Han, T.~Li and T.~Plehn,
  %``Measuring CP Violating Phases at a Future Linear Collider,''
  Phys.\ Lett.\  B {\bf 475}, 342 (2000).
%  [arXiv:hep-ph/9907425].
  %%CITATION = PHLTA,B475,342;%%



%\cite{Hill:1983xh}
\bibitem{Hill:1983xh}
  C.~T.~Hill,
  %``Are There Significant Gravitational Corrections To The Unification
  %Scale?,''
  Phys.\ Lett.\  B {\bf 135}, 47 (1984).
  %%CITATION = PHLTA,B135,47;%%


%\cite{Shafi:1983gz}
\bibitem{Shafi:1983gz}
  Q.~Shafi and C.~Wetterich,
  %``Modification Of GUT Predictions In The Presence Of Spontaneous
  %Compactification,''
  Phys.\ Rev.\ Lett.\  {\bf 52}, 875 (1984).
  %%CITATION = PRLTA,52,875;%%


%\cite{Ellis:1984bm}
\bibitem{Ellis:1984bm}
  J.~R.~Ellis, C.~Kounnas and D.~V.~Nanopoulos,
  %``No Scale Supersymmetric Guts,''
  Nucl.\ Phys.\  B {\bf 247}, 373 (1984).
  %%CITATION = NUPHA,B247,373;%%

%\cite{Ellis:1985jn}
\bibitem{Ellis:1985jn}
  J.~R.~Ellis, K.~Enqvist, D.~V.~Nanopoulos and K.~Tamvakis,
  %``Gaugino Masses And Grand Unification,''
  Phys.\ Lett.\  B {\bf 155}, 381 (1985).
  %%CITATION = PHLTA,B155,381;%%

%\cite{Drees:1985bx}
\bibitem{Drees:1985bx}
  M.~Drees,
  %``Phenomenological Consequences Of N=1 Supergravity Theories With Nonminimal
  %Kinetic Energy Terms For Vector Superfields,''
  Phys.\ Lett.\  B {\bf 158}, 409 (1985).
  %%CITATION = PHLTA,B158,409;%%

%\cite{Anderson:1999uia}
\bibitem{Anderson:1999uia}
  G.~Anderson, H.~Baer, C.~h.~Chen and X.~Tata,
  %``The reach of Fermilab Tevatron upgrades for SU(5) supergravity models  with
  %non-universal gaugino masses,''
  Phys.\ Rev.\  D {\bf 61}, 095005 (2000).
%  [arXiv:hep-ph/9903370].
  %%CITATION = PHRVA,D61,095005;%%


%\cite{Chamoun:2001in}
\bibitem{Chamoun:2001in}
  N.~Chamoun, C.~S.~Huang, C.~Liu and X.~H.~Wu,
  %``Non-universal gaugino masses in supersymmetric SO(10),''
  Nucl.\ Phys.\  B {\bf 624}, 81 (2002).
%  [arXiv:hep-ph/0110332].
  %%CITATION = NUPHA,B624,81;%%


%\cite{Chakrabortty:2008zk}
\bibitem{Chakrabortty:2008zk}
  J.~Chakrabortty and A.~Raychaudhuri,
  %``A note on dimension-5 operators in GUTs and their impact,''
  Phys.\ Lett.\  B {\bf 673}, 57 (2009).
%  [arXiv:0812.2783 [hep-ph]].
  %%CITATION = PHLTA,B673,57;%%


%\cite{Martin:2009ad}
\bibitem{Martin:2009ad}
  S.~P.~Martin,
  %``Non-universal gaugino masses from non-singlet F-terms in non-minimal
  %unified models,''
  Phys.\ Rev.\  D {\bf 79}, 095019 (2009).
%  [arXiv:0903.3568 [hep-ph]].
  %%CITATION = PHRVA,D79,095019;%%

%\cite{Bhattacharya:2009wv}
\bibitem{Bhattacharya:2009wv}
  S.~Bhattacharya and J.~Chakrabortty,
  %``Gaugino mass non-universality in an SO(10) supersymmetric Grand Unified
  %Theory: low-energy spectra and collider signals,''
  Phys.\ Rev.\  D {\bf 81}, 015007 (2010).
%  [arXiv:0903.4196 [hep-ph]].
  %%CITATION = PHRVA,D81,015007;%%

%\cite{Feldman:2009zc}
\bibitem{Feldman:2009zc}
  D.~Feldman, Z.~Liu and P.~Nath,
  %``Gluino NLSP, Dark Matter via Gluino Coannihilation, and LHC Signatures,''
  Phys.\ Rev.\  D {\bf 80}, 015007 (2009).
%  [arXiv:0905.1148 [hep-ph]].
  %%CITATION = PHRVA,D80,015007;%%


%\cite{Chamoun:2009nd}
\bibitem{Chamoun:2009nd}
  N.~Chamoun, C.~S.~Huang, C.~Liu and X.~H.~Wu,
  %``Intermediate Scale Dependence of Non-Universal Gaugino Masses in
  %Supersymmetric SO(10),''
  arXiv:0909.2374 [hep-ph].
  %%CITATION = ARXIV:0909.2374;%%

%\cite{Calmet:2008df}
\bibitem{Calmet:2008df}
  X.~Calmet, S.~D.~H.~Hsu and D.~Reeb,
  %``Grand unification and enhanced quantum gravitational effects,''
  Phys.\ Rev.\ Lett.\  {\bf 101}, 171802 (2008).
%  [arXiv:0805.0145 [hep-ph]].
  %%CITATION = PRLTA,101,171802;%%



%%%%%%%%%%%%%%%%%%%%%%%%%%%%%%%%%%%%%%%%%%%%%%%%%%%%%%%%%%%%%%%%%%%%%%
%     F-Theory References
%%%%%%%%%%%%%%%%%%%%%%%%%%%%%%%%%%%%%%%%%%%%%%%%%%%%%%%%%%%%%%%%%%%%%%

%\cite{Vafa:1996xn}
\bibitem{Vafa:1996xn}
  C.~Vafa,
  %``Evidence for F-Theory,''
  Nucl.\ Phys.\  B {\bf 469}, 403 (1996).
%  [arXiv:hep-th/9602022].
  %%CITATION = NUPHA,B469,403;%%


%\cite{Donagi:2008ca}
\bibitem{Donagi:2008ca}
  R.~Donagi and M.~Wijnholt,
  %``Model Building with F-Theory,''
  arXiv:0802.2969 [hep-th].
  %%CITATION = ARXIV:0802.2969;%%

%\cite{Beasley:2008dc}
\bibitem{Beasley:2008dc}
  C.~Beasley, J.~J.~Heckman and C.~Vafa,
  %``GUTs and Exceptional Branes in F-theory - I,''
  JHEP {\bf 0901}, 058 (2009).
%  [arXiv:0802.3391 [hep-th]].
  %%CITATION = JHEPA,0901,058;%%

%\cite{Beasley:2008kw}
\bibitem{Beasley:2008kw}
  C.~Beasley, J.~J.~Heckman and C.~Vafa,
  %``GUTs and Exceptional Branes in F-theory - II: Experimental Predictions,''
  JHEP {\bf 0901}, 059 (2009).
%  [arXiv:0806.0102 [hep-th]].
  %%CITATION = JHEPA,0901,059;%%


%\cite{Donagi:2008kj}
\bibitem{Donagi:2008kj}
  R.~Donagi and M.~Wijnholt,
  %``Breaking GUT Groups in F-Theory,''
  arXiv:0808.2223 [hep-th].
  %%CITATION = ARXIV:0808.2223;%%




%\cite{Font:2008id}
\bibitem{Font:2008id}
  A.~Font and L.~E.~Ibanez,
  %``Yukawa Structure from U(1) Fluxes in F-theory Grand Unification,''
  JHEP {\bf 0902}, 016 (2009).
%  [arXiv:0811.2157 [hep-th]].
  %%CITATION = JHEPA,0902,016;%%





%\cite{Jiang:2009zza}
\bibitem{Jiang:2009zza}
  J.~Jiang, T.~Li, D.~V.~Nanopoulos and D.~Xie,
  %``F-SU(5),''
  Phys.\ Lett.\  B {\bf 677}, 322 (2009).
  %%CITATION = PHLTA,B677,322;%%


%\cite{Blumenhagen:2008aw}
\bibitem{Blumenhagen:2008aw}
  R.~Blumenhagen,
  %``Gauge Coupling Unification In F-Theory Grand Unified Theories,''
  Phys.\ Rev.\ Lett.\  {\bf 102}, 071601 (2009).
%  [arXiv:0812.0248 [hep-th]].
  %%CITATION = PRLTA,102,071601;%%






%%%%%%%%%%%%%%%%%%%%%%%%%%%%%%%%%%%%%%%%%%%%%%%%%%%%%%%%%%%%%%%%%%%%%%%%%%%%


%\cite{Jiang:2009za}
\bibitem{Jiang:2009za}
  J.~Jiang, T.~Li, D.~V.~Nanopoulos and D.~Xie,
  %``Flipped SU(5) X U(1)_X Models from F-Theory,''
  Nucl.\ Phys.\  B {\bf 830}, 195 (2010).
%  [arXiv:0905.3394 [hep-th]].
  %%CITATION = NUPHA,B830,195;%%




%%%%%%%%%%%%%%%%%%%%%%%%%%%%%%%%%%%%%%%%%%%%%%%%%%%%%%%%%%%%%%%%%%%%%%%%%%%%

%\cite{Li:2009cy}
\bibitem{Li:2009cy}
  T.~Li,
  %``SU(5) and SO(10) Models from F-Theory with Natural Yukawa Couplings,''
  arXiv:0905.4563 [hep-th].
  %%CITATION = ARXIV:0905.4563;%%


%%%%%%%%%%%%%%%%%%%%%%%%%%%%%%%%%%%%%%%%%%%%%%%%%%%%%%%%%%%%%%%%%%%%%%
%%%%%%%%%%%%%%%%%%%%%%%%%%%%%%%%%%%%%%%%%%%%%%%%%%%%%%%%%%%%%%%%%%%%%%

%\cite{Leontaris:2009wi}
\bibitem{Leontaris:2009wi}
  G.~K.~Leontaris and N.~D.~Tracas,
  %``Gauge coupling flux thresholds, exotic matter and the unification scale in
  %F-SU(5) GUT,''
  arXiv:0912.1557 [hep-ph].
  %%CITATION = ARXIV:0912.1557;%%



%%%%%%%%%%%%%%%%%%%%%%%%%%%%%%%%%%%%%%%%%%%%%%%%%%%%%%%%%%%%%%%%%%%%%%
%%%%%%%%%%%%%%%%%%%%%%%%%%%%%%%%%%%%%%%%%%%%%%%%%%%%%%%%%%%%%%%%%%%%%%



%\cite{Li:2010mr}
\bibitem{Li:2010mr}
  T.~Li, J.~A.~Maxin and D.~V.~Nanopoulos,
  %``F-Theory Grand Unification at the Colliders,''
  arXiv:1002.1031 [hep-ph].
  %%CITATION = ARXIV:1002.1031;%%


%%%%%%%%%%%%%%%%%%%%%%%%%%%%%%%%%%%%%%%%%%%%%%%%%%%%%%%%%%%%%%%%%%%%%%%%%%%%

%%%%%%%%%%%%%%%%%%%%%%%%%%%%%%%%%%%%%%%%%%%%%%%%%%%%%%%%%%%%%%%%%%%%%%

%%%%%%%%%%%%%%%%%%%%%%%%%%%%%%%%%%%%%%%%%%%%%%%%%%%%%%%%%%%%%%%%%%%%%%
\bibitem{TLDN-P}
  T.~Li and D.~V.~Nanopoulos, in preparation.

%%%%%%%%%%%%%%%%%%%%%%%%%%%%%%%%%%%%%%%%%%%%%%%%%%%%%%%%%%%%%%%%%%%%%%

%%%%%%%%%%%%%%%%%%%%%%%%%%%%%%%%%%%%%%%%%%%%%%%%%%%%%%%%%%%%%%%%%%%%%%


%\cite{Georgi:1979dq}
\bibitem{Georgi:1979dq}
  H.~Georgi and D.~V.~Nanopoulos,
  %``Ordinary Predictions From Grand Principles: T Quark Mass In O(10),''
  Nucl.\ Phys.\  B {\bf 155}, 52 (1979).
  %%CITATION = NUPHA,B155,52;%%


%%%%%%%%%%%%%%%%%%%%%%%%%%%%%%%%%%%%%%%%%%%%%%%%%%%%%%%%%%%%%%%%%%%%%%

%%%%%%%%%%%%%%%%%%%%%%%%%%%%%%%%%%%%%%%%%%%%%%%%%%%%%%%%%%%%%%%%%%%%%%



%%%%%%%%%%%%%%%%%%%%%%%%%%%%%%%%%%%%%%%%%%%%%%%%%%%%%%%%%%%%%%%%%%%%%%

%%%%%%%%%%%%%%%%%%%%%%%%%%%%%%%%%%%%%%%%%%%%%%%%%%%%%%%%%%%%%%%%%%%%%%



\end{thebibliography}
\end{document}